\documentclass[11pt,a4paper]{article}
\pdfoutput=1
\usepackage{jheppub}
\usepackage{amsmath,amscd}
\usepackage{amsfonts}
\usepackage{amssymb}
\usepackage{graphicx}
\usepackage{color}
\usepackage[normalem]{ulem}
\usepackage{hyperref}
\usepackage{enumerate}
\newcommand{\CC}{\mathbb{C}} % Complessi
\newcommand{\RR}{\mathbb{R}} % Reali
\newcommand{\ZZ}{\mathbb{Z}} % Interi
\newcommand{\NN}{\mathbb{N}} % Naturali
\def\tr         {{\rm  tr}}

\def\calf         {{\cal F}}

\def\calh         {{\cal H}}

\def\call         {{\cal L}}
\def\calm         {{\cal M}}

\def\calo         {{\cal O}}

\def\cals         {{\cal S}}

\def\be{\begin{equation}}
\def\ee{\end{equation}}
\def\bea{\begin{eqnarray}}
\def\eea{\end{eqnarray}}

\def\a{\alpha}
\def\b{\beta}
\def\h{\eta}
\def\g{\gamma}

\def\d{\delta}
\def\e{\epsilon}

\def\f{\phi}

\def\n{\nu}
\def\o{\omega}

\def\p{\pi}
\def\r{\rho}

\def\s{\sigma}

\def\t{\tau}

\def\sF{{{ F}\!\!\!\!\hskip.8pt\hbox{\raise1pt\hbox{/}}\,}}
\def\som{{{ \omega}\!\!\!\!\hskip.8pt\hbox{\raise1pt\hbox{/}}\,}}
\def\sJ{{{\rm J}\!\!\!\!\hskip.8pt\hbox{\raise1pt\hbox{/}}\,}}

\def\F{\Phi}
\def\pa{\partial}

\def\to{\rightarrow}
\def\nonu{\nonumber \\{}}
\def\half{{1 \over 2}}

\title{Conical spaces, modular invariance and $c_{p,1}$ holography}
\author[a]{Joris Raeymaekers}

\affiliation[a]{CEICO, Institute of Physics of the Czech Academy of Sciences,\\  Na Slovance 2, 182 21 Prague 8, Czech Republic.}
\emailAdd{joris@fzu.cz}

\abstract{We propose a non-unitary example of  holography for the family of two-dimensional logarithmic conformal field theories 
	 with  negative central charge $c= c_{p,1} = - 6p +13 - 6 p^{-1}$.  We argue that at large $p$, these models have a semiclassical gravity-like description which  contains, besides the global AdS$_3$ spacetime, a tower of solitonic solutions describing conical excess angles. Evidence comes from the fact that the central charge and the natural modular invariant partition function of such a theory  coincide with those of
the $c_{p,1}$ model.
These theories  have an extended chiral W-algebra whose  currents have  large spin of order $|c|$, and which in the bulk are realized as spinning conical solutions. 
As a by-product we also find a direct link between geometric actions for  exceptional Virasoro coadjoint orbits, which describe fluctuations around the conical spaces, and Felder's free field construction of degenerate representations.}
\arxivnumber{}
\begin{document}
 \maketitle
\section{Introduction}
In the search for simple models of quantum gravity, it is natural to attempt  to give a holographic description of pure gravity  in three-dimensional anti-de-Sitter (AdS) spacetime. Such a two-dimensional holographic dual conformal theory (CFTs) is governed by Virasoro symmetry and should  satisfy   basic consistency conditions such as crossing symmetry and modular invariance.

In addition, one would ideally like to impose further requirements in order to obtain a realistic theory of quantum gravity in the bulk, such as
\begin{enumerate}[I]
 \item The existence of a large central charge limit in which the bulk gravity theory becomes semiclassical. 
\label{req1}
    \item The requirement that  conformally invariant vacuum, dual to the unperturbed global AdS spacetime, is a normalizable state in the spectrum.\label{req2}
    \item Unitarity.\label{req3}
\end{enumerate}
However, the natural guess for the modular invariant  partition function under these assumptions, due to Maloney and Witten \cite{Maloney:2007ud}
and refined more recently in  \cite{Benjamin:2019stq,Benjamin:2020mfz,Maxfield:2020ale}, appears not to describe a single dual theory but rather an ensemble average over  CFTs.
Improving our understanding of such an `imprecise holography'
is an active area of current research which ties in with recent insights into the gravitational path integral and the black hole information puzzle.

In view of these conceptual issues 
it is of interest to take a step back and relax some of the conditions  \ref{req1}-\ref{req3}, at the cost of making the theory less realistic. Several such examples have appeared in the literature. One instance which does not obey  \ref{req1} is Witten's proposal of the `monster' CFT being dual to pure gravity\footnote{Another  example is the proposed dual  gravity dual interpretation of the unitary minimal models with $0<c<1$ \cite{Castro:2011zq,Jian:2019ubz}.} at $c=24$ \cite{Witten:2007kt}. If one is willing to give up requirement \ref{req2}, a natural proposal is to consider Liouville theory at large $c$, in  which  the conformally invariant vacuum is not part of the spectrum. This idea goes back to \cite{Coussaert:1995zp} at the classical level and was proposed to extend to the quantum level in \cite{Li:2019mwb}.

In this work, we  propose a holographic duality 
where we instead drop the requirement  \ref{req3} of unitarity. In particular we will consider CFTs at large but negative central charge. A hint that CFTs at large but  negative central charge may be under better control than their counterparts at positive $c$ comes from the fact that in  $(p,p')$ minimal models the central charge is given
by
\be 
c_{p,p'} = 1 - 6 {(p -p')^2 \over  p p'},\label{cMM}
\ee
 and can be made arbitrary large and negative\footnote{The large negative $c$ limit of the $(2 p- 1,2)$ minimal models, when coupled to gravity \cite{Anninos:2020geh} or as part of minimal string theory \cite{Saad:2019lba}, has been interpreted as a semiclassical bulk theory with a matrix integral description, in the latter case related to JT gravity.}. 

On the bulk side, negative central charge corresponds to  taking Newton's constant to be negative. Although gravity in three dimensions does not have any propagating degrees of freedom, the change of sign does influence the response in the presence of sources. While a point mass  source leads to conical  deficit at positive $c$ \cite{Deser:1983tn,Deser:1983nh}, it leads to a conical surplus in the case of  negative $c$.  
In fact, one encounters a special case of (\ref{cMM})  when considering  certain spaces with a conical surplus and their holographic interpretation. In \cite{Castro:2011iw},  conical excess angles which are an integer multiple $r$ of  $2\p $  (as well as their higher spin generalizations) were shown to be regular solitons in the Chern-Simons description of the theory. 
These have conformal weight
\be 
h^{(r)} = - { (r^2-1) c \over 24} + \calo (c^0 ).
\ee
and form, at negative $c$,  tower of states above the global  AdS solution with $r=1$.
These solutions possess extra symmetries which, upon quantization, correspond to the presence of a null vector \cite{Perlmutter:2012ds,Raeymaekers:2014kea}.
In \cite{Campoleoni:2010zq}, it was argued that  the compatibility of this symmetry with conformal invariance requires the central charge to be a special case of (\ref{cMM}),
 namely $c_{|k|,1}$, where the (negative) integer $k$ is the quantized level of the $sl(2,\RR) \times sl(2,\RR)$ Chern-Simons description.
 The limiting   cases of (\ref{cMM}) with $p'=1$ actually don't correspond to  Virasoro
  minimal models, but there do exist  well-studied $c_{|k|,1}$ logarithmic CFTs. They  have an extended chiral algebra containing a triplet of spin-$(2 |k| - 1)$ currents (see \cite{Flohr:2001zs,Gaberdiel:2001tr} for reviews). These models  are rational in the sense that they contain a finite number of representations of the extended algebra.
  
   In this work we extend these ideas into a proposal for a holographic duality for the logarithmic $ c_{|k|,1}$ CFTs at large $p$.  One of the main observations  is that the spectrum of   the $c_{|k|,1}$ logarithmic CFTs arises as the natural modular invariant extension of the pure gravity  spectrum at negative $k$. 
   Logarithmic CFTs have made an appearance in the holographic correspondence in the description of 3D gravity at the chiral point  (see \cite{Grumiller:2013at}  and references therein) and, more generally in the   holographic description of singletons \cite{Kogan:1999bn}.

 The structure of the paper is as follows.  The first part of the paper is dedicated to clarifying the description of quantized surplus solutions in terms of  chiral bosons, extending the results of \cite{Campoleoni:2017xyl}. 
 We begin by reviewing the role of conical surplus spaces as solitons in the Chern-Simons formulation of gravity in section \ref{Secconexc}, and generalize the results of \cite{Cotler:2018zff} to show that fluctuations around these solutions are captured by the geometric action for exceptional orbits of the Virasoro group.  In section \ref{Secfreefield} we introduce  a field redefinition which maps this action to that of a free chiral boson \cite{Floreanini:1987as}, albeit with an extra constraint. Imposing this constraint  in the quantum theory while preserving conformal invariance leads \cite{Campoleoni:2017xyl} to  the value $c_{|k|,1}$ for the central charge. We also find a direct link between our description and Felder's construction \cite{Felder:1988zp} of degenerate representations as subspaces of free field Fock spaces.

In section \ref{Secmodinv}  we address the issue of embedding the  states found by quantizing Lorentzian gravity solutions  into a consistent modular invariant CFT. Since the surplus solutions live in  winding sectors around the spatial circle in the free boson description, it is natural to extend the Euclidean path integral on the torus to include also the winding sectors along the Euclidean time direction. One is then naturally led to the partition function of a  compact boson  at radius $R = \sqrt{|k|/2}$.  This is indeed the partition function for the logarithmic $c_{|k|,1}$ models \cite{Flohr:1995ea,Gaberdiel:1996np} and neatly  decomposes  into degenerate characters of both the Virasoro algebra \cite{DiFrancesco:1987gwq} and the extended triplet W-algebra. An interesting feature is that the W-currents, whose spin scales with $|c|$, are in this case realized as solitonic bound states of the gravitational field, rather than as additional gauge fields in the bulk.  We end with a comment on averaged holography and list some confusions and possible generalizations.

\section{AdS$_3$ gravity, conical spaces and exceptional orbits}\label{Secconexc}
Classical Einstein gravity in three spacetime dimensions permits an equivalent formulation  as a Chern-Simons theory. While at the fully nonperturbative  quantum level the relation between the two formulations remains unclear,  the Chern-Simons formulation at least seems to give a sensible perturbative description around a given background. In this section we review the Chern-Simons description and single out a class of smooth solitonic solutions. These correspond to conical excess angles in the metric formulation. We show that, when expanded around these solutions, the Chern-Simons action reduces  to the geometric action for the exceptional orbits of the Virasoro group.
\subsection{Gravity in Chern-Simons variables}
Three dimensional AdS gravity can be reformulated as a Chern-Simons theory with action \cite{Achucarro:1987vz,Witten:1988hc}
\be
S = S [A] - S [\tilde A], \qquad 
 S [A] = {k \over 4\p} \tr  \int \left( A  \wedge dA + {2 \over 3} A \wedge A \wedge A \right). \label{CSaction}
\ee
 Our choice of gauge group is $PSL(2,\RR)\times \widetilde{PSL(2,\RR)}$. The  group $PSL(2,\RR) \sim SL(2,\RR)/\ZZ_2$ consists of real $2\times 2$ matrices with unit determinant, modulo
 the equivalence relation
 \be
 g \sim - g.
 \ee
 This $\ZZ_2$ quotient is  important for global considerations: as it will turn out, is needed in order to allow the global AdS$_3$ background as a nonsingular Chern-Simons configuration \cite{Castro:2011iw}. As explained in  \cite{Witten:2007kt}, for the Chern-Simons path integral to be well-defined, the level $k$ has to be quantized to be a multiple of 4:
 \be 
 k \in 4 \ZZ .\ee
 The fact that $k$ is an integer will play an important role in what follows. 
 %for a comparison of various global gauge groups in 3D gravity).
 The Lie algebra is
 \be
 [J_a, J_b  ]= \e_{ab}^{\ \ c} J_c,
 \ee
 where indices are raised with $\h_{ab}= {\rm diag} (-1,1,1)$.
 We will also use the $L_m, m = 0, \pm 1$ basis
 \be 
 L_0 = - J_2, \qquad L_{\pm 1}= J_0 \pm J_1.
 \ee
 For definiteness, we take the trace in (\ref{CSaction}) is taken in the 2-dimensional  defining representation where the generators can be taken to be
 \be 
 J_0 = {i \over 2} \s_2, \qquad J_1 = -{1 \over 2} \s_1, \qquad J_2 = -{1 \over 2} \s_3.\label{matrixrepr}
 \ee

 The Chern-Simons formulation is related to the standard formulation of gravity in terms of  vielbein and spin connection variables through
 \be
 A^a = \o^a + {1 \over l} e^a, \qquad \tilde A^a = \o^a - {1 \over l} e^a.\label{eomdef}
 \ee
 Upon substituting in (\ref{CSaction}) one obtains, up to a boundary term, the Einstein-Hilbert action\footnote{Parity invariance results from initially considering coordinate systems where  $\det e = \sqrt{ - g} >0$ and then declaring  that $A$ and $\tilde A$ are exchanged under parity, so that  that $e$  is parity-odd. }:
 \be 
  S =  {k \over 4 \p l}\int_\calm \det e \left(  R + {2 \over l^2} \right) - {k \over 4 \p }\int_{\d \calm} \o^a e_a.
 \ee
 A Chern-Simons configuration can be only be interpreted as a regular gravity solution if the vielbein is invertible,  
 \be \det ( A - \tilde A)  \neq 0.
 \ee
The level $k$ in the Chern-Simons description is related to the Brown-Henneaux central charge \cite{Brown:1986nw} $c = 3l/2G$ as
 \be 
c =  6k+ \calo\left( k^{0} \right).\label{cclass}
 \ee
Here, we indicated that this is a classical relation valid at large $k$ and that there could be quantum corrections.

For the solutions we will be interested in,  the manifold $\calm$ has the topology of a solid cylinder $\RR \times D$, and impose boundary conditions on the spatial boundary
 \be 
 A_- = \tilde A_+ =0 \qquad {\rm on \ } \RR \times S^1,\label{bc}
 \ee
 where
 \be 
 x_\pm \equiv \f \pm t.
 \ee
 One checks that with these boundary conditions the variational principle based on (\ref{CSaction}) is well-defined.
 
 \subsection{WZW form of the action}
 The Chern-Simons action is  first order in time derivatives, and 
to clarify the  canonical structure of the theory it is useful to perform a 1+2 split
 \bea
 d &=& dt { \pa \over \pa t} +\hat d, \\ 
 A &=& A_t dt + \hat A,
 \eea
to find
 \be
  S[A]  = -{k \over 4\p} \tr \int_{\RR \times D} dt \wedge \left(  \hat A  \wedge\dot{\hat A} - 2 A_t \hat F \right)-  {k \over 4\p} \int_{\RR \times S^1} dt \wedge d\f  \tr(  A_t \hat A_\f ).
 \ee
 In the boundary term, we can replace $ \tr(  A_t \hat A_\f )$ by 
 $ \tr A_\f^2$ by adding a term which vanishes under our boundary conditions (\ref{bc}). Similar operations on $S[\tilde A]$ lead to
the following   first order action compatible with the boundary conditions (\ref{bc}) 
 \bea
 S  &=&-{k \over 4\p} \tr \int_{\RR \times D} dt \wedge \left(  \hat A \wedge \dot{\hat A}-  2 A_t \hat F \right) +  {k \over 4\p} \tr \int_{\RR \times D} dt \wedge\left(  \hat {\tilde A}\wedge  \dot{\hat {\tilde A}}-  2 \tilde A_t \hat{\tilde F} \right)\nonu
 && - {k \over 4\p} \tr \int_{\RR \times S^1} dt d\f (\hat A_\f^2+ \hat { \tilde A}_\f^2 ).
 \eea
 The fields $A_t$ and $\tilde A_t$ are Lagrange multipliers enforcing the constraints
 \be
 \hat F = \hat{\tilde F} =0.
 \ee
 These  are solved by setting
 \be 
 \hat A = G^{-1} \hat d G, \qquad \hat{\tilde A} =\tilde G^{-1} \hat d \tilde G.
 \ee
 Here, $G(t, \r, \f)$ and $ \tilde G(t, \r, \f)$ are spacetime dependent $PSL(2,\RR)$ group elements and $\r$ denotes the radial coordinate transverse to the boundary.

 In terms of these variables the action becomes
 \bea 
 S &=&  {k \over 2 \p} \tr\int_{\RR \times S^1}dt \wedge d\f \pa_\f G^{-1}\pa_- G + {k \over 12 \p} \tr \int_{\RR \times D} (G^{-1} d G)^3\nonu
 &+&  {k \over 2 \p} \tr\int_{\RR \times S^1}dt\wedge d\f \pa_\f\tilde G^{-1}\pa_+\tilde G - {k \over 12 \p} \tr \int_{\RR \times D} (\tilde G^{-1} d\tilde G)^3.\label{chiralWZW}
 \eea
  in which we recognize the difference of two chiral  WZW actions \cite{Sonnenschein:1988ug} on $PSL(2, \RR)$. The equations of motion following from  (\ref{chiralWZW}) are
 \be
 \pa_\f ( G^{-1}\pa_- G ) = \pa_\f ( \tilde G^{-1}\pa_+\tilde G)=0.\label{eomWZW}
 \ee

In what follows we will be interested in smooth gauge fields $\hat A$ and $\hat{\tilde A}$, which in particular have trivial holonomy around the contractible $\f$-circle.  This requires that the group elements $G(t, \r, \f)$ and $ \tilde G(t, \r, \f)$  are single-valued,
\be 
G(\f+ 2 \p ) = \pm G( \f), \qquad  \tilde G(\f+ 2 \p ) = \pm \tilde G( \f).\label{smoothG}
\ee
 
 \subsection{Asymptotic conditions}
 Next we impose that the fields behave near the boundary  like the global AdS$_3$ solution, where
 \be 
 G_{AdS} = e^{ - \half x_+ (L_1 + L_{-1})} e^{ \r L_0}, \qquad \tilde G_{AdS} = e^{ - \half x_- (L_1 + L_{-1})} e^{ - \r L_0}.\label{AdSsol}
 \ee
 Concretely we take this to mean (see \cite{Campoleoni:2010zq} for a detailed justification)  that the group elements factorize,
 \be 
 G = g( t, \f) e^{ \r L_0}, \qquad \tilde G = \tilde  g( t, \f) e^{ - \r L_0},\label{raddecomp}
 \ee
 while the elements $g$ and $\tilde g$, which live on the boundary cylinder, satisfy so-called  Drinfeld-Sokolov \cite{Drinfeld:1984qv} constraints:
 \be
 \tr L_{-1} g^{-1} \pa_\f g = {\theta \over 2}, \qquad \tr L_{1}\tilde  g^{-1} \pa_\f \tilde g = {\tilde \theta \over 2}, \qquad {\rm where \ } \theta, \tilde \theta > 0.
 \label{DS}
 \ee
 All positive values of $ \theta, \tilde \theta $ are equivalent as they can be rescaled by sending $g \to g e^{a L_0}$. 

 \subsection{A history of excess}
 The group element $g (t, \f)$ is a map from the boundary cylinder into $PSL(2,\RR)$. Since $PSL(2,\RR)$ is not simply connected, such maps  come in distinct topological classes \cite{Elitzur:1989nr} that will play an important role in this work. 
  Indeed, the topology of $PSL(2,\RR)$ is $ S^1\times \RR^2 $,
 where the $S^1$ corresponds to the 
 maximal compact subgroup SO(2). This can be seen more explicitly by using the Iwasawa decomposition, `$g = KAN$':
 \bea 
  g &=& e^{- \F J_0}  e^{2R L_0} e^{   H L_{-1}}\\
  &=& \left(\begin{array}{cc}  \cos {\F \over 2}  &- \sin {\F \over 2} \\  \sin {\F \over 2} &\cos {\F \over 2} \end{array} \right)
 \left(\begin{array}{cc} e^{R} &0\\ 0 &e^{-R} \end{array} \right) \left(\begin{array}{cc} 1 & H \\ 0 &1 \end{array} \right).
 \label{Iwasawa}
 \eea
  and similarly for $\tilde g$. In the second line we have used the matrix representation (\ref{matrixrepr}).
 The noncompact  coset space swept out by $AN$ has trivial $\RR^2$ topology while the compact element $K$ describes an $S^1$. The normalization of $\F$ is chosen such that the $PSL(2, \RR)$  identification   $g \sim -g$ implies the periodicity $\F \sim \F + 2 \p$.
 The map $g (t, \f)$ 
 is therefore characterized by a winding number counting how many times the $\f$-circle direction of the cylinder is wound around the $S^1$ parametrized by $\F$. In these winding sectors the fields  $\F, \tilde \F$ satisfy:
 \be 
 \F (t, \f + 2 \p) = \F + 2 \p r,\qquad \tilde \F (t, \f + 2 \p) =\tilde  \F + 2 \p \tilde r .\label{Phiwind}
 \ee
 A representative solution of the equations of motion (\ref{eomWZW}) in the sector with winding numbers $r$ and $\tilde r$ is of the form (\ref{raddecomp}) with
 \be 
 \F = r x_+, \qquad  \tilde \F = \tilde r x_-, \label{surplPhi}
 \ee
 and $R= \tilde R = H =\tilde H=0$.
  For strictly positive $r$ and $\tilde r$ this solution  also satisfies the asymptotic conditions  (\ref{DS}), with $ \theta=r,  \tilde \theta=\tilde r$. The group elements  are single-valued (\ref{smoothG})  and therefore lead to smooth Chern-Simons gauge fields with trivial holonomy \cite{Castro:2011iw}.
 
 As we see from (\ref{AdSsol}), the global AdS background belongs to the class of winding solutions with $r = \tilde r =1$\footnote{The fact that the $AdS$ group element behaves as $g (\f + 2 \p) = -g$ was the reason for the $\ZZ_2$ quotient in our choice of the global gauge group $SL(2,\RR)/\ZZ_2$ \cite{Castro:2011iw}.}, and most discussions of 3D gravity in Chern-Simons variables restrict the solution space to this sector. 
 Upon quantization, the state space one obtains is the  Virasoro vacuum representation, as can be seen from several points of view (see e.g. \cite{Maloney:2007ud,Cotler:2018zff}). 
 
 In this work we will  instead consider general smooth solutions obeying the asymptotic conditions and therefore to include the other $(r, \tilde r)$ sectors as well\footnote{A different space of solutions which has been explored in the literature is to instead restrict to Chern-Simons gauge fields with hyperbolic holonomy. As we have just seen, this excludes the global AdS vacuum solution. This prescription naturally leads to Liouville theory \cite{Coussaert:1995zp,Li:2019mwb} which indeed doesn't contain the conformally invariant vacuum as a normalizable state.}. The energy of these solutions, computed from (\ref{chiralWZW}), is
 \be 
 H = - {k (r^2 + \tilde r^2) \over 4}.
 \ee
 Therefore, if we want to retain  all winding while keeping the  energy bounded from below, we  should confine ourselves to the regime of negative level (and central charge)
 \be 
 k <0,
 \ee
 which we will assume from now on. 
 
Let us briefly review the geometry of these solutions. For the solutions (\ref{surplPhi}) with $r= \tilde r >1$ the metric computed from  (\ref{eomdef}) is
 \be
 ds^2 = l^2 \left(- \cosh^2 \r d(rt)^2 + d\r^2 + r^2 \sinh^2 \r d\f^2 \right).\label{surplmetr}
 \ee
This represents static a spacetime with a  `conical surplus' singularity  in the origin $\r = 0$, where there is an angular excess  of $2 \p (r-1)$. For $r \neq \tilde r$ on obtains a spinning generalization of this metric, and
  we will loosely refer to all the solutions with general winding numbers as `surpluses'. The  reason that these smooth Chern-Simons  gauge fields lead to metrically singular spaces  is that the vielbein computed from them degenerates at the origin in a way that, unlike the case $r=\tilde r=1$, is not just a coordinate singularity.

 One would expect that  quantization of the more general $r, \tilde r$ winding sectors leads to other Virasoro representations besides the vacuum module. Indeed, there is by now strong evidence \cite{Perlmutter:2012ds,Campoleoni:2013iha,Raeymaekers:2014kea,Campoleoni:2017xyl}
 that the relevant representation is $(r, 1) \otimes (\tilde r, 1)$, where we denote by $(r,s)$ the degenerate representations in the Kac classification \cite{Kac:1978ge}. We will review extend these arguments in section \ref{Secfreefield}. A similar interpretation was argued to hold for higher spin generalizations of the surplus solutions. The possible relevance of winding sectors in the Chern-Simons formulation was mentioned in \cite{Witten:2007kt}, and early work discussing the surpluses appears in \cite{Izquierdo:1994jz,Mansson:2000sj,deBoer:2014sna}.

 Before continuing we should mention another geometry which will make a somewhat unexpected appearance. This is the zero mass and zero angular momentum limit of the BTZ black hole metric, which arises as an $r\to 0$ limit of (\ref{surplmetr}):
 \be 
 ds^2 = l^2 \left( d\r^2 + e^{2\r} ( -dt^2 + d\f^2 ) \right). \label{BTZM0}
 \ee
The corresponding  group elements  are
\be 
g = e^{- \half {x_+} L_1}, \qquad  \tilde g = e^{- \half {x_-} L_{-1}}.
\ee 
These are not single valued and lead to a holonomy of the gauge connections which is of parabolic type. This solution has energy $H=0$ and lies below global AdS in our regime of  negative $k$. 
At present there is no  clear reason to include this solution in the theory, but  in section \ref{Secmodinv} we will find that the  modular invariant completion of the theory does contain  states with   energy $H = \calo (k^0)$.
 
 \subsection{Surpluses and exceptional Virasoro orbits}\label{Secexc}
 After this digression on the space of solutions we are interested in  we continue the reduction the theory
 under the asymptotic constraints (\ref{DS}). 
 We will do this at the level of the action (\ref{chiralWZW}), generalizing the discussion for the vacuum sector in \cite{Cotler:2018zff}. This will give a new perspective on  the relation between the winding sectors (\ref{Phiwind}) and exceptional coadjoint orbits of the Virasoro group, which was discussed from the perspective of the equations of motion in \cite{Raeymaekers:2014kea}. 
 
 As the above discussion suggests, it is for our purposes  convenient to parametrize the group element   using the Iwasawa decomposition (\ref{Iwasawa}).
 In
 this parametrization 
 the left-invariant one-form is
  \be 
 g^{-1} dg = - {e^{2 R} \over 2} d\F L_1 + \left(2 dR - He^{2 R} d\F \right)L_0 + \left( 2 H dR + dH - \half\left( e^{-2 R} + H^2 e^{2 R}\right) d\F \right) L_{-1},\label{leftinv}
 \ee
 and working out the Lagrangian density in the first line of (\ref{chiralWZW})  one finds
 \be 
 \call_L = {|k| \over 2 \p} \left( 2  R' \pa_- R - \half \F' \pa_- \F - e^{2 R}  \F' \pa_- H \right),\label{LLR}
 \ee
 where a prime denotes a derivative with respect to $\f$.
 
 Imposing the asymptotic condition (\ref{DS}) we can eliminate $R$,
 \be 
 e^{- 2 R} = { \F'\over \theta }, \label{asR}
 \ee
 and, recalling that $\theta >0$, this imposes in addition that $\F$ is monotonic,
 \be 
 \F'>0.\label{Phimon}
 \ee
Substituting (\ref{asR}) in the Lagrangian density (\ref{LLR}),  we obtain
 \be 
 \call_L =  {|k| \over 4 \p}\left( {\F'' \pa_- \F'\over (\F')^2} - \F'\pa_- \F - \theta \pa_- H \right).\label{LPhi}
 \ee
 We note that the field $H$  enters only through a total derivative and can be dropped. 
 The fact that $H$ disappears from the dynamics
  reflects the invariance of the asymptotic conditions (\ref{DS}) under the `Drinfeld-Sokolov' gauge transformation \cite{Drinfeld:1984qv}
 \be 
 g \to g \left(\begin{array}{cc} 1 & \a \\ 0 &1 \end{array} \right),
 \ee 
 under which $H$ gets shifted, $H \to H + \a$.

 The dynamics described by (\ref{LPhi}) is closely related to the theory of coadjoint orbits of the Virasoro group \cite{Witten:1987ty}. Each winding sector labelled by $r$ captures the effect of adding boundary gravitons to the corresponding surplus solution (\ref{surplPhi}), and one would expect that it describes a particular coadjoint orbit.  This was made precise for the $r=1$ sector by Cotler and Jensen \cite{Cotler:2018zff}, who pointed out that
 the action (\ref{LPhi}) is essentially the Alekseev-Shatashvili geometric action \cite{Alekseev:1988ce} on the  coadjoint orbit of  the vacuum.
 
 To generalize this to the other winding sectors, we recall that the Virasoro group is a central extension of the group of diffeomorphisms of the circle. 
 The latter can be represented as smooth functions $\Psi (\f)$
 satisfying
 \be 
\Psi'>0, \qquad \Psi (\f + 2 \p) =    \Psi ( \f) + 2 \p. \label{diffS1}
 \ee
 A geometric action, which defines a symplectic structure equivalent to the Kirillov-Kostant symplectic structure on coadjoint orbits,  
 can be formulated in terms of a field variable $\Psi (t, \f)$ satisfying (\ref{diffS1}) at all times.
 
 Focusing on our theory (\ref{LPhi}) in the sector with winding number $r$, where $\F$ satisfies $\F ( \f + 2 \p) = \F ( \f)  + 2 \p r$, the rescaled field 
 \be \Psi  \equiv {\F \over r} \ee
 satisfies (\ref{diffS1}) and describes a diffeomorphism of the circle.
 The Lagrangian (\ref{LPhi}) reads
 \be
 \call_L^{(r)} =  {|k| \over 4 \p}\left( {\Psi'' \pa_- \Psi'\over (\Psi')^2} + B_r \Psi'\pa_- \Psi \right),\label{LAS}
 \ee
 with 
 \be 
B_r = - r^2. \label{b0exc}
 \ee
 In (\ref{LAS})  we recognize the geometric action on the coadjoint orbit with constant representative $B$  \cite{Alekseev:1988ce}. More precisely, as we shall see shortly, it is a chiral version of the action presented in \cite{Alekseev:1988ce}.
  The specific value of $B_r$ arising in (\ref{b0exc}) for our description of the surpluses are very special: it corresponds to the so-called $r$-th exceptional orbit which possess extra symmetries compared to a generic orbit. As we will review below, these symmetries are related to a null vector appearing at level $r$ upon quantization, generalizing the null vector from acting with $L_{-1}$   in the case of the vacuum module.

 Let us review some of the properties of this geometric action, reverting to the description\footnote{To verify  some of the formulas below, it is useful to make field redefinition  $
 	F = e^{i \F}$
 	in terms of which the action simplifies,
 	\be 
 	\call_L = {k\over 4 \p} {F'' \pa_- F'\over (F')^2} + {\rm \ total \ derivatives}.
 	\ee} (\ref{LPhi}) in terms of $\F$.
To see the appearance of a chiral Virasoro symmetry, let us find the variation of the action under an arbitrary reparametrization of the $\f$-coordinate:
 \be 
 \f \to \f - \e (t, \f), \qquad \d \F = - \e (t, \f) \F'.
 \ee
 One finds that
 \be 
 \d \call_L = - {1 \over \p} \pa_- \e \, T + {\rm derivative\ terms,}
 \ee
 where 
  \be 
T =- {|k| \over 2}\left( \cals (\F,\f ) + \half (\F')^2 \right),\label{TPhi}
 \ee
 and where $\cals (F,\f )$ is the Schwarzian derivative
 \be
 \cals (\F,\f ) \equiv {\F'''\over \F' } - {3 \over 2} \left( {\F''\over \F' }\right)^2.
 \ee
 Therefore  the  time component of the $\f$-translation current, which we denote by $T$, is purely left-moving   on-shell
\be 
 \pa_- T = 0,
 \ee
and this  is precisely the equation of motion following from (\ref{LPhi}). 
The conserved $\f$-momentum is
\be 
J = { 1 \over 2 \p} \int_0^{2\p} d\f\, T = - H
\ee
This dispersion relation is typical of a chiral theory.
To see  the second equality, we observe that $-T/2 \p$ is equal to the canonical Hamiltonian density plus an improvement term
\be
-{T \over 2 \p} = \calh_{\rm can}+{|k| \over 4 \p} \left( {\F''\over \F'} \right)'.
\ee
In addition to $H = -J \equiv L_0 -c/24$, the theory has conserved Virasoro currents
\be 
L_n = -{ 1 \over 2 \p} \int_0^{2\p} d\f\, e^{i n x_+} T +{ c \over 24} \d_{n,0}.\label{LsVir}
\ee

 An important property of the action (\ref{LPhi}) describing exceptional Virasoro orbits is that it is invariant under  a group $PSU(1,1) \sim PSL(2,\RR)$ of restricted gauge transformations. Under these, $e^{i \F}$ transforms by a time-dependent fractional linear transformation,
 \be 
e^{i \F} \to {a(t) e^{i \F} + b(t)\over \bar b(t) e^{i \F} + \bar a (t) }, \qquad |a|^2 - |b|^2 =1,\label{gaugePhi}
 \ee
 or, at the infinitesimal level,
 \be 
 \d \F = \e_0(t) + \e (t) e^{i \F} +  \bar \e (t) e^{- i \F}  .\label{gaugePhiinf}
 \ee

 Another property which we should point out is that all the exceptional orbits  contain tachyonic directions in our regime of negative $k$.
 Indeed, Fourier decomposing the field   in the $r$-th 
 winding sector,
 \be
 \F = r \f +  \sum_{m \in \ZZ} \g_m e^{i m \f}, 
 \ee 
 the energy becomes
 \be 
 H = {|k| r^2 \over 4} + {|k| \over  2} \sum_{m \in \NN} m^2 \left(1- {m^2\over r^2}\right) \g_m \g_{-m} + \ldots \label{tachmodes}
 \ee
Therefore the modes  with $m>r$ are tachyonic\footnote{By contrast, when $k$ is positive the tachyonic modes are those with $m<r$ and are therefore absent in the vacuum sector.}.  We note that the flat directions $m= 0, \pm r$ correspond to the gauge transformations (\ref{gaugePhiinf}). 
 
 In general, the quantization of the Kirillov-Costant Poisson bracket on a Virasoro coadjoint orbit yields a representation of the Virasoro group. Using the geometric action, this translates into the statement that the Euclidean   path integral of the theory on the torus should yield a Virasoro character. In the  case at hand however, the path integral over real $\F$ is ill-defined due to the tachyonic modes in (\ref{tachmodes}).  The standard remedy \cite{Gibbons:1978ac}  is to  
  analytically continue the integration contour such that $H$ is bounded from below. From (\ref{tachmodes}),  the normal modes $\b_m$ are
 \be 
 \b_m = - i \sqrt{|k|\over 2} m \left( 1+ {m \over r} \right) \g_m + \calo(|k|^0),\label{betas}
 \ee
 and the energy is bounded below if we take the reality condition on the $\b_m$ to be
 \be 
 \b_m^* = \b_{-m}, \label{betacontour}
 \ee 
 which does not correspond to a real field $\F$.
The Euclidean path integral on the torus along the contour (\ref{betacontour}) can be performed and generalizes of the calculation for the $r=1$ case in \cite{Cotler:2018zff}.  We shall perform this  calculation in section \ref{SecEuclPI} below using  different variables which are more convenient for our purposes. As might be expected the result, cfr. (\ref{c1loop},\ref{h1loop}),  yields the character of a  non-unitary, lowest weight, representation of the Virasoro algebra.

	\section{Free field variables and the Coulomb gas}\label{Secfreefield}
In this section we discuss the reformulation of the geometric actions on  exceptional Virasoro orbits (\ref{LAS}) in terms of free field variables. This description is essentially a Lagrangian version of the variables introduced in \cite{Campoleoni:2017xyl}\footnote{The main difference with that work is that we work with the Lorentzian gauge group $PSL(2,\RR)^2$ instead of $SL(2,\CC)$. In the Lorentzian case, the gauge group is not simply connected and requires the quantization of the level $k$ which in turn will  lead to the central charge $c_{p,1}$ with $p= |k|$.}. It will clarify the connection to the Coulomb gas formalism \cite{Dotsenko:1984nm}  and, in particular, to Felder's  construction of irreducible Virasoro representations as subspaces of free field Fock spaces \cite{Felder:1988zp} (see \cite{Kapec:2020xaj} for a modern perspective).
 \subsection{It's a chiral boson, Jim, but not as we know it}
 We start by making a field redefinition to a field $X$ satisfying
 \be
e^{i X} = ( e^{ i \F} )'.\label{XitoPhi}
 \ee 
 It follows from the periodicity of $\F$ that $X$ is again a compact direction with period $2\p$.  
 The Lagrangian density (\ref{LPhi}) becomes, up to boundary terms\footnote{More precisely, the Lagrangians are related as
 \be 
 \call_L'= \call_L + {ik \over 4\p}\left( \pa_- \left( 2 \F'+ { \F \F'' \over \F'} \right) - \pa_\f \left( { \F \pa_- \F' \over \F'}\right)\right).
 \ee}
 , quadratic  in the field $X$:
 \be 
 \call_L' = - {k \over 4 \p} X'\pa_- X.\label{XLag}
 \ee
 This is in fact the standard  Lagrangian  for  a free chiral boson due to  Floreanini and Jackiw  \cite{Floreanini:1987as}. The equation of motion is
 \be 
 \pa_\f (\pa_- X ) =0.\label{eomX}
 \ee
 The action has a restricted gauge symmetry descending from the transformation (\ref{gaugePhiinf}) with parameter $\e_0$, under which
 \be 
 X \to X + \e_0 (t).\label{shiftX}
 \ee
 This can be used to select  purely leftmoving solutions to (\ref{eomX}) where $\pa_- X =0$.
 
 Several comments and refinements  concerning  this field redefinition are in order, though. First of all, we should note that under (\ref{XitoPhi}), the real field $\F$  is in general not mapped to a real field $X$. Nevertheless, the surplus solutions of interest stay (up to an irrelevant constant) real and become winding solutions of $X$,
 \be 
 X  = r x_+.
 \ee
 In fact, expanding $X$ in normal modes at $t=0$ as 
 \be 
 X = r \f - i \sqrt{2 \over k} \sum_{m \neq 0} {\a_m \over m} e^{-im\f},
 \ee
 we see from (\ref{XitoPhi}) that the $\a_m$  for $m \neq \pm r$ are, to leading order at large  $|k|$, precisely the normal modes (\ref{betas})  in the geometric theory:
 \be
 \a_m = \b_m + \calo \left(|k|^{-\half} \right), \qquad {\rm for\ } m \neq \pm r.\label{bitoa}
 \ee
 Therefore, the path integral over a real field $X$ with $\a_m^* = \a_{-m}$ corresponds precisely to the  analytic continuation (\ref{betacontour}) required to make original the path integral over $\F$ well-defined.

 Secondly, as it stands 
  the free theory (\ref{XLag}) is not equivalent to the geometric action (\ref{LPhi}) since it describes an enlarged phase space.  Indeed, we already saw in (\ref{bitoa}) that the modes $\a_{\pm r}$ do not have an equivalent in the original description. The reason for this is that the expression for $\F$ in terms of $X$ is nonlocal,
\be 
e^{i \F(t,\f)} = \int_{\f_0}^\f e^{i X (t, \f ')} d \f',\label{PhiitoX}
\ee
with $\f_0$ an arbitrary integration constant.
In the original theory, the function $e^{i \F(t,\f)}$ must be periodic in $\f$,  but under (\ref{PhiitoX}) this will not be the case for a generic solution $X$ of the equation (\ref{eomX}).
It is easy to see that periodicity of $e^{i \F}$ imposes a further constraint on $X$,
\be 
Q_- \equiv \int_0^{2 \p} d \f e^{ i X (t,\f)} =0.\label{Xconstr}
\ee

Imposing  (\ref{Xconstr}) is also necessary for 
the action  to possess
the full $PSL(2, \RR)$ gauge symmetry  (\ref{gaugePhiinf}) without  which, as was stressed in \cite{Cotler:2018zff}, the theory would not capture the correct physics.
 The  $PSL(2, \RR)$ gauge transformations parametrized by $\e$ in (\ref{gaugePhiinf})  act nonlocally on $X$ as
 \be
 \d X =  \e (t) \int_{\f_0}^\f e^{i X(t,\f')} d\f' + \bar \e (t) \int_{\f_0}^\f e^{-i X(t,\f')} d\f'.\label{gaugeX}
 \ee
  Focusing for the moment on the $\e$ transformation, one finds that
for constant $\e$, the action is formally invariant due to the conservation of the nonlocal  Noether current
\be 
(j_\e^t, j_\e^\f) = \left( e^{i X}, - e^{i X} + 2 i \pa_- X  \int_{\f_0}^\f e^{i X(t,\f')} d\f'\right).
\ee
The corresponding Noether charge is however local and given by the quantity $Q_-$ defined in  (\ref{Xconstr}). To make the theory invariant also for time-dependent $\e$, we should gauge this global symmetry which amounts to imposing the constraint (\ref{Xconstr}).

 At the level of the action, we impose the constraint (\ref{Xconstr}) by introducing a Lagrange multiplier field $C_t(t)$ depending only on time and modify the Lagrangian to
\be 
\call'_L= - {|k| \over 4\p} \left( X'\pa_- X + i C_t e^{iX} - i \bar C_t e^{- iX} \right).\label{XLagfinal}
\ee
The action becomes formally\footnote{By this, we mean that it transforms up to boundary terms which vanish when (\ref{Xconstr}) holds.}  gauge invariant under (\ref{gaugeX}) with $A_t$ transforming a s gauge potential,
\be 
 \d C_t (t)= \dot \e (t) - i \e_0 (t) C_t(t).\label{gaugeC}
 \ee
The second term is needed for the new terms in the action to preserve the shift symmetry (\ref{shiftX}) with parameter $\e_0$. The equations of motion following from (\ref{XLagfinal}) read
 \bea
\pa_- X'- \half \left(C_t e^{i X} + \bar C_t e^{-iX}\right) &=&0\\
\int_0^{2\p} d\f e^{i X(t,\f)} &=&0.\label{eomfinal}
\eea

To have a good variational principle, the above Lagrangian has to be supplemented by appropriate boundary conditions on the fields and possible boundary terms. We have already  imposed quasi-periodic boundary conditions on $X$ when going around the $\f$-circle, but  we should also specify boundary conditions at initial and final times. As was stressed in \cite{Henneaux:1987hz}, for an action which is first-order in time derivatives one cannot impose that the fields are fixed both at initial and final times. The required boundary term, which will not play a role in what follows, is reviewed in Appendix \ref{Appbdyterm}.

  The stress tensor (\ref{TPhi}) becomes
 \be 
T =- {|k| \over 2} \left(\half (X')^2+i  X''\right),  \label{TX}
 \ee
which is the standard free field stress tensor with an improvement term indicating the presence of a background charge which shifts the value of the central charge. 
 The stress tensor (\ref{TX}) is invariant under the gauge transformations (\ref{gaugeX}). In fact, the gauge invariance (\ref{shiftX},\ref{gaugeX}) can also be derived as the most general transformations leaving the stress tensor (\ref{TX}) invariant \cite{Campoleoni:2017xyl}.
 Without gauging $Q_-$, the free field theory (\ref{XLag})  would therefore describe many copies of the same exceptional coadjoint orbit.
 
 Before continuing our discussion of the formulation (\ref{XLagfinal}),  we would like to connect it to the earlier work \cite{Campoleoni:2017xyl} on free field variables for gravity and higher spin theories. The discussion there was phrased in terms of the Chern-Simons gauge fields and the starting point was a specific gauge choice for the Drinfeld-Sokolov transformations (\ref{DS}). These can be used to make the $L_{-1}$-component of the connection vanish; this is called the `diagonal gauge' and was first discussed in \cite{Balog:1990mu}. In our parametrization (\ref{leftinv}), the diagonal gauge imposes a first order differential equation on the field $H$,
	\be 
\theta H^2 - 2 H'+ 2 {\F'' \over \F'} H  =0.\label{diaggauge}
\ee
A  specific solution is 
\be 
H = { i \over \theta} \F',
\ee
and using the relation (\ref{XitoPhi}) between $\F$ and $X$, the gauge field takes the form
\be
a_\f= g^{-1} g' = -{\theta \over 2} L_1 - i X'L_0.\label{adiag}
\ee
	This is precisely  the parametrization of the diagonal gauge used  in \cite{Campoleoni:2017xyl}.
However, the diagonal gauge condition (\ref{diaggauge}) does not fix $H$ completely, but leaves an arbitrary time-dependent integration constant. It's straightforward to see that the effect of turning on  this integration constant, while preserving the form (\ref{adiag}), is that $X$ transforms precisely as under the $PSL(2,\RR)$ transformation  (\ref{gaugeX}). Therefore this part of the $PSL(2,\RR)$ symmetry of the system  can also be viewed as the residual gauge freedom  \cite{Campoleoni:2017xyl} left over after imposing the diagonal gauge.

 In summary, we propose  the theory based on  (\ref{XLagfinal}) in the sector with winding number $r$ as an alternative to the  geometric action (\ref{LPhi}) for the $r$-th exceptional coadjoint orbit. As we already mentioned, this orbit was argued to lead, upon quantization, to the degenerate Virasoro representation of type $(r,1)$ in Kac's classification. We will present two further arguments in favour of this identification, which will also serve as a check on our proposed description  (\ref{XLagfinal}). Firstly, we will see that the Euclidean  path integral of (\ref{XLagfinal}) on the torus yields the character of the $(r,1)$ representation, and, secondly, that  the quantization of 
 (\ref{XLagfinal}) in the operator formalism  leads to a standard construction \cite{Felder:1988zp} of the irreducible $(r,1)$ representation as a subspace of a free field Fock space.

 \subsection{Euclidean path integral on the torus}\label{SecEuclPI}
 In this subsection we will calculate, as promised at the end of section \ref{Secexc}, the  Euclidean path integral of the theory (\ref{XLagfinal}) on a torus with complex structure modulus $\t$. This should compute a  trace
 \be 
 Z= \tr_\calf e^{ - 2 \p \t_2 H - 2 \p  i \t_1 P}=  \tr_\calf  q^{ L_0 - {c \over 24}},
 \ee
over the quantum state space $\calf$. Here  we used that $P=-H\equiv-(L_0-{c \over 24})$, see (\ref{LsVir}), and defined $ q \equiv e^{2\p i \t}$.
 One therefore expects that 
for each coadjoint orbit characterized by  winding number $r$, the path integral should yield the character of an irreducible Virasoro representation.

We start by analytically continuing the  action (\ref{LPhi}) to Euclidean signature, $t =  i t_E,\  S_E= i S$, and find
 \be 
 S_E = {|k| \over 4\p} \int dt_E d\f \left( X'\pa_{\bar w} X + C e^{i X} - \bar C e^{ - i X}\right),
 \ee
 where $w \equiv \f + i t_E$. 
  Next, we make an additonal  identification such that the theory lives on a torus with modular parameter $\t= \t_1 + i \t_2$:
 \be 
 (t_E, \f) \sim (t_E, \f+ 2\p)\sim (t_E + 2 \p \t_2, \f+ 2\p \t_1).
 \ee
 The boundary condition appropriate for $r$-th surplus solution is
 \bea 
 X^{(r)} (t_E, \f+ 2\p) &=& X^{(r)} (t_E, \f) + 2 \p r\\
 X^{(r)} (t_E + 2 \p \t_2, \f+ 2\p \t_1) &=& X^{(r)} (t_E, \f)\\
 C^{(r)} (t_E + 2 \p \t_2) &=&  C^{(r)} (t_E).
 \eea
 The classical solution obeying these boundary conditions is (up to gauge transformations of the form (\ref{gaugeX})),
 \be 
 X_{\rm cl}^{(r)} = r \left( \f - {\t_1 \over \t_2} t_E\right),\qquad C_{\rm cl}^{(r)} =0 .
 \ee
 We work at large $|k|$ and expand the fields in modes around the classical solution
 \bea 
 X^{(r)} &=& X_{\rm cl}^{(r)} + {1 \over \sqrt{|k|}} \sum_{m,n \in \ZZ} a_{m,n}
 e^{i \left( m \f + {n- m \t_1\over \t_2}t_E \right) },\label{Xexp}\\
 C^{(r)} &=& C_{cl}^{(r)} +  {1 \over \sqrt{|k|} \t_2} \sum_{n\in \ZZ} b_n  e^{ in t_E \over \t_2},
 \eea 
 the   action becomes, to quadratic order in the fluctuations, 
 \be 
 S_E^{(r)} = - 2\pi i \t {|k| r^2\over 4} - {i \p   } \sum_{m,n \in \ZZ}  m  (m  \t -n)  |a_{m,n}|^2   + i \p \sum_{n \in \ZZ} \left( b_n  \bar a_{r,n}   + \bar   b_n  a_{r,-n} + \calo (|k|^{-\half}) \right)  
 . \label{Squadr}
 \ee
 We note that, at this order, the role of the gauge field is simply to set to zero the superfluous modes $a_{\pm r,n}$, which as we saw in (\ref{bitoa}) do not correspond to fluctuations in the original theory (\ref{LPhi}). At higher orders though, the integral over the gauge field will introduce interaction terms.

 To one-loop order, the  integral over the modes\footnote{The attentive reader will notice that, following \cite{Cotler:2018zff}, we have judiciously chosen the $\t_2$-dependence of the measure so that the result depends holomorphically on $\t$.} $a_{m,n},b_n$ leads to
 \be
 Z_{\rm (1-loop)}^{(r)} =  q^{|k| r^2 \over 4} \prod_{m \in \ZZ,m\neq 0,\pm  r}^\infty \prod_{n \in \ZZ} (- i m(n - m  \t))^{-\half}.
\ee
Differentiating  with respect to $\t$, the sum over $n$ converges for each $m$ and leads to (ignoring $\t$-independent normalization factors)
\bea
Z_{\rm (1-loop)}^{(r)}  &=&
  q^{{|k| r^2 \over 4} - {r \over 2}+ \half \sum_{m=1}^\infty m}{(1- q^r)\over   \prod_{m=1}^\infty  \left(1- q^{{m}}\right)} .
 \eea
Using  zeta function regularization to interpret the infinite sum as $
 \sum_{m=1}^\infty m= - {1 \over 12} $,
 we obtain the result
 \be 
 Z_{\rm (1-loop)}^{(r)}  = {q^{ {|k| r^2 \over 4} - {r \over 2}}(1- q^r)\over \h},
 \label{Z1loop} \ee
 where 
  $\h = q^{1\over 24}\prod_{m=1}^\infty (1- q^m)$ is  Dedekind's eta function.
 
 From this expression we can read off the 1-loop correction to the central charge and the energy of the surplus solutions. For $r=1$,
 (\ref{Z1loop}) has the form of the vacuum character $ \chi_0 =  q^{- {c-1 \over 24}}(1-q)/ \h$ in a CFT with
 \be 
 c= - 6 |k| + 13 + \calo (|k|^{-1}).\label{c1loop}
 \ee
 The shift by 13 reproduces the result found in \cite{Cotler:2018zff} using the original variables (\ref{LPhi}).
 
 For $r > 1$,  the result (\ref{Z1loop}) is the character for a primary representation with weight $h^{(r)}-{c / 24}={|k| r^2 / 4} - {r / 2} - {1 / 24}+ \calo (|k|^{-1})$, or, expressed in terms of $c$, 
 \be 
 h^{(r)}-{c \over 24}
 = - {c r^2 \over 24} + {(r-1) (13 r + 1) \over 24} + \calo (c^{-1} ).\label{h1loop}
 \ee
The second term in brackets in (\ref{Z1loop}) means that there is a null vector at level $r$.  One checks that the 1-loop correction term in (\ref{h1loop})  agrees with the large-$c$ limit of the primary weight in the $(r,s=1)$ degenerate representation
   in Kac's classification  \cite{Kac:1978ge} (see (\ref{hrs}) below for the all-order expression). This representation  indeed has a null vector at level $r$.
 
  The above computation 
   gives a path  integral version of the proposed  quantization  of the exceptional orbits in \cite{Raeymaekers:2014kea}, where the same 1-loop correction (\ref{h1loop}) was found from the operator quantization of the natural Poisson bracket on the exceptional orbit  in a large $c$ expansion.

 \subsection{Relation to Coulomb gas and Felder's construction}
 In this section, we will consider  quantization of the model (\ref{XLagfinal}) in the operator formalism. It consists of the standard  chiral boson action supplemented with an extra  constraint $Q_- =0$ which corresponds to a nonlocal gauge symmetry (\ref{gaugeX}).
 Since it is not clear how to extend the standard  methods of constrained quantization to nonlocal symmetries, our approach will be to first quantize the model without the  constraint and then to impose it as an operator equation on the resulting state space.  This approach was also used in other examples with nonlocal gauge symmetry \cite{Lavelle:1993xf},  and in the present context it leads to results in agreement with the path integral calculation in the previous section.
It  also gives a natural derivation of the all-order corrected versions of (\ref{c1loop}) and (\ref{h1loop}), and provides a direct link with the free field  construction of the $(r,1)$ representations.

The first step is the quantization of the Floreanini-Jackiw Lagrangian (\ref{XLag}). 
This has been discussed in detail in \cite{Henneaux:1987hz,Sonnenschein:1988ug} which we now briefly review.
 The theory has a constraint \be P = {k \over 8\p} X', \label{constrXP}\ee
 as well as a restricted gauge invariance (\ref{shiftX}). The latter is fixed by restricting the space of solutions to purely leftmoving functions 
 \be 
 \dot X = X'.\label{leftm}
 \ee
 The outcome of the Dirac bracket analysis is
 \be
 \{ X(t,\f) , P (t, \f') \}_{DB} = {k \over 8\p} \{ X(t,\f) , X' (t, \f') \}_{DB} = \half \d(\f- \f'),\label{DBX}
 \ee
 which differs by a factor$1/2$ from the naive Poisson bracket. A useful check at this point is that the Dirac bracket $ \{ X(\f) , X ( \f') \}_{DB} \sim {\rm sgn} (\f- \f')$ implies that the charge $Q_-$ defined in (\ref{Xconstr}) is indeed the canonical generator of the nonlocal transformation (\ref{gaugeX}). 
 
 Upon expanding $X$ in modes 
 \be 
 X = x +  {2  \over |k|} p_{\rm cyl} x_+ - i \sqrt{2 \over |k|}\sum_{m \in \ZZ_{\neq 0}} {\a_m \over m}e^{- i m x_+},
 \ee
 and quantizing the Dirac bracket, we find the canonical commutation relations
 \be
 [x, p_{\rm cyl}] = {i}, \qquad [\a_m, \a_{n} ]=  m \d_{m+ n,0}.\label{CCR}
 \ee
 
We now discuss the quantization of eigenvalues of $ p_{\rm cyl}$. If we view $X$ as the leftmoving part of a nonchiral boson, $ p_{\rm cyl}$ is quantized as
\be 
p_{\rm cyl} =\half (r |k| - s), \qquad r,s \in \ZZ,\label{quantizedmom}
\ee
where $r$ is, as before, the winding number and  $s$ is a momentum quantum number. If we were studying a pure chiral boson theory, the constraint (\ref{leftm}) would impose the vanishing of the rightmoving momentum \cite{Gross:1985fr,Henneaux:1987hz}  and impose $s = - r |k|$.
However, in the present context, this is too restrictive. One reason is that, due to the  background charge and the resulting anomalous transformation law of the momentum current, this relation would not be preserved when mapping the cylinder to the plane.  
Therefore we allow the more general quantization condition (\ref{quantizedmom}) appropriate for the chiral part of a non-chiral boson.

 In view of this remark we define ground states labelled (redundantly)  by winding $r$ and momentum  $s$  which satisfy
 \be 
  p_{\rm cyl} |r,s\rangle =  \half (r |k| - s) |r,s\rangle, \qquad  \a_{m>0}  |r,s\rangle =0, \qquad  |r+1,s+|k|\rangle =  |r,s\rangle.
  \ee
  We denote the Fock space built on $|r,s\rangle$ with the creation modes
  as $\calf_{r,s}$.
The Laurent expansion on the Euclidean plane reads, using
 $t = i t_E, x_+ = w, x_- = \bar w, z = e^{ i w}$.
\be 
X = x - i {2  \over |k|} p \ln z - i \sqrt{2 \over |k|}\sum_{m \in \ZZ_{\neq 0}} {\a_m \over m z^m},
\ee
leading to the basic OPE 
\be
\pa X (z) \pa X(0) \sim - {2 \over |k| z^2}.
\ee
The quantum stress tensor is
\be 
T = - {|k| \over 4} : (\pa X)^2 : - {i \over 2} (|k| + q_0) \pa^2 X,\label{Tquantq0}
\ee
where we have allowed an order one correction $q_0$ to the background charge, for reasons to become clear presently. The conformal weight of an exponential operator 
\be 
V_p = :e^{i p X}: \label{expvo}
\ee
is then
\be
h_p = { p (p + |k| + q_0) \over |k|}.\label{weightexp}
\ee

 As we mentioned above, we want to furthermore  impose the constraint $Q_-=0$ on the quantum state space. Since $Q_-$ is a composite operator we have to give an ordering prescription to
  define it unambiguously. 
A  natural choice is to define $Q_-$ using conformal normal ordering, 
\be 
Q_- = \oint dz : e^{i X} (z) :.
\ee
As already noted in \cite{Campoleoni:2017xyl}, the compatibility of the
projection on states with $Q_-=0$  with conformal invariance requires a quantum shift of the backgound charge, i.e. turning on  $q_0$ in (\ref{Tquantq0}).
Indeed, in order for $Q_-$ to  commute with the Virasoro generators,  the primary operator $ : e^{i X}:$ should have weight one\footnote{A similar argument, going back to \cite{Curtright:1982gt}, determines the quantum shift of the background charge in Liouville theory by requiring that the Liouville potential is a weight $(1,1)$ primary.}.  
 From (\ref{weightexp}) this fixes 
$q_0 = -1$ and the  quantum stress tensor is
\be 
T = - {|k| \over 4} : (\pa X)^2 : - {i \over 2} (|k|-1) \pa^2 X.\label{Tquant}
\ee
The quantum correction to the background charge leads to the corrected central charge
\be 
c = - 6 |k|  + 13 - { 6 \over |k|}. \label{cquant}
\ee
The  term of order one agrees with the 1-loop computation (\ref{c1loop}).
Recalling the expression (\ref{cMM}) for the central charge $c_{p,p'}$ in the $(p,p')$ minimal model
we see that (\ref{cquant}) corresponds to
\be 
c = c_{|k|,1}.
\ee

Because of the presence of a background charge in the stress tensor,
the momentum current $j\equiv {i k\pa X/2}$ transforms anomalously under conformal tranformations $z \to \tilde z$:%. 
\be 
\pa_z \tilde z j_{\tilde z} (\tilde z) = j_z (z) -{|k|-1\over 2} {\pa^2_z \tilde z \over \pa_z \tilde z}.
\ee
The momentum quantum numbers on the cylinder and the plane are therefore related as
\be 
p = p_{\rm cyl} - {|k|-1\over 2}.
\ee
As a result, the state-operator mapping for the ground states  is
\be 
|r,s\rangle\qquad  \longrightarrow \qquad V_{r,s} \equiv : e^{{i \over 2} \left((r-1) |k| - (s-1) \right)X}(0):\label{stateop}
\ee
and their conformal weight is, from (\ref{weightexp}),
\be 
h_{r,s} = {(|k| r - s)^2 \over 4|k|} + {c-1 \over 24}.\label{hrs}
\ee
When $rs$ is strictly positive, $h_{r,s}$ is precisely the primary weight of  the degenerate  representation  with Kac labels $(r,s)$\footnote{The fact that the degenerate representations can be realized as momentum and winding states of a compact boson was pointed out in \cite{Felder:1988zp}.}. The Fock space $\calf_{r,s}$ forms a reducible Virasoro module with a null vector   
 at level $rs$ \cite{Feigin:1982tg}.  
 
Let us return to the surplus solutions, which as we saw from the classical solution, live in  winding sectors.
Our path integral computation\footnote{The present discussion suggest that  the exact result for the torus path integral contains an additional 2-loop contribution $Z_{\rm exact }^{(r)} =  q^{1 \over 4 k} Z_{\rm (1-loop)}^{(r)}= \chi_{r,1}$, where $Z^{\rm (1-loop)}$ is given in (\ref{h1loop}). 
	The 2-loop contribution to the central charge (\ref{cquant}) 
	seems somewhat at odds  
	with the argument of \cite{Cotler:2018zff}, using localization techniques, that the $r=1$ path integral is in fact 1-loop exact.
	We interpret the discrepancy as indicating that our  $r=1$ path integral is obtained in a renormalization scheme differing by a 2-loop counterterm. It appears that this is the scheme which yields results in agreement with the present operator formalism using conformal normal ordering  for composite operators such as $:e^{i X}:$. It would be interesting to understand this point better. %as 
} in the previous section (cfr.  (\ref{h1loop})) 
identified them with  the primary states $|r,s=1 \rangle$, and therefore we should assign to them one unit of momentum on the cylinder. This was not visible at the classical level and is a consequence of the background charge. Without this assignment, the global AdS background in the $r=1$ sector would not correspond to the identity operator and would not preserve the AdS symmetries. Similar objections would hold for  the $r>1$ sectors.

In summary,  the quantum fluctuations around the surplus of winding number $r$ live in the Fock space $\calf_{r,1}$. The second  step in our construction is to project on the states satisfying $ Q_-=0$. First  we note that a general vertex operator $V_{r,s}$ in  (\ref{stateop}) can be seen as a map between the Fock spaces
\be 
V_{r,s}: \calf_{r',s'} \to \calf_{r'+ r-1,s' + s-1}.
\ee
Since $Q_-$ can be written as
\be 
Q_- = \oint dz V_{1,-1} (z),
\ee
it maps
\be 
Q_-: \qquad \calf_{r,1} \to \calf_{r,-1}.
\ee
The state space $\calh_{r,1}$ we  are interested in is the kernel of this map, 
 \be 
\calh_{r,1} \equiv  \ker_{\calf_{r,1}} Q_-.
 \ee 
This space is precisely the  irreducible $(r,1)$ representation as can be seen as follows. Since the image of  $Q_-$ lies in $\calf_{r,-1}$ and we have
 \be 
h_{r,-1}  = h_{r,1} + r,
\ee 
one can show that the  ${\rm im}\  Q_-$ is precisely the Virasoro submodule formed by the null vector at level $r$ and its descendants \cite{Felder:1988zp}. The  states  in $\ker Q_-$ therefore span the  irreducible $(r,1)$ module.  This free field realization is precisely  Felder's construction \cite{Felder:1988zp} of the $(r,1)$ modules at  central charge\footnote{For the Virasoro minimal model central charges $c_{p,p'}$ with $1<p'<p$ the construction  is more involved and reduces to a cohomology problem.} (\ref{cquant}) (see also \cite{Flohr:1995ea}).

For later reference, we collect here some facts about the free field realization of the  irreducible $(r,s)$ representations  with $s>1$, which will play a role in what follows.
Because of the identities 
\be 
h_{r,s} = h_{r+1, s+ |k|} = h_{-r,-s},
\ee
we can restrict $s$ to the range
\be 
1 \leq s \leq |k|,
\ee
Defining the $s$-th power of $Q_-$ with a specific contour prescription as in \cite{Felder:1988zp},
the operator $Q_-^s$ maps
 \be 
Q_-^s: \qquad \calf_{r,s} \to \calf_{r,-s},
\ee
and using the fact that 
\be 
h_{r,-s}  = h_{r,s}+ rs
\ee 
one can again  argue \cite{Felder:1988zp} that
\be 
\calh_{r,s} \equiv \ker_{\calf_{r,s}} Q_-^s
\ee
carries the irreducible $(r,s)$ representation.
The $(r,s)$ character is
\be 
\chi_{r,s} = \tr_{\calh_{r,s}} q^{L_0 -{c \over 24}} = { q^{(|k| r - s)^2\over 4 |k|}\over \h} (1 - q^{rs}),\label{charvir}
\ee
and these degenerate representations obey  abstract fusion rules of the form \cite{Flohr:1995ea}
\be
\calo_{r_1,s_1} \times  \calo_{r_2,s_2} = \sum_{\footnotesize\begin{array}{c} r_3 = |r_1-r_2|\\ r_1+ r_2+r_3=1  \ {\rm mod}\  2 \end{array}}^{r_1+r_2}\sum_{\footnotesize \begin{array}{c} s_3 = |s_1-s_2|\\ s_1+ s_2+s_3=1 \ {\rm mod}\  2 \end{array}}^{\min (|k|,s_1+s_2-1) } \calo_{r_3,s_3}.\label{fusion}
\ee

\section{Modular invariant extension and $c_{|k|,1}$ models}\label{Secmodinv}
In the previous sections we studied smooth Lorentzian solutions and their  boundary graviton excitations and found that, upon quantization,  give rise to the $(r,1)$ degenerate representations of the Virasoro algebra. We now address the issue of assembling these representations into a modular invariant theory. This will require extending the state space to include also the more general $(r,s>1)$ representations. These appear rather naturally in the path integral on the torus from including winding sectors around the Euclidean time direction. Doing this we will  led to a spectrum matching that of the  logarithmic triplet CFT at $c_{|k|,1}$. This theory has an extended chiral symmetry algebra with W-algebra currents of large spin $2|k|-1$, which in the bulk are realized as spinning surplus solutions.
 \subsection{Surpluses and the triplet algebra}
The arguments of the previous section suggest that the state space of pure gravity in the regime of large negative central charge includes degenerate modules of the type \be\calh_{r,1} \otimes \calh_{\tilde r,1}\ee at central charge $c_{|k|,1}$, at least for some\footnote{ Not all values of $r$ and $\tilde r$
	need to appear, for example to impose invariance under the modular $T$ transformation it is natural to 
	rule out those values of $r$ and $\tilde r$  for which the difference $h_{r,1}-  h_{\tilde r,1}$ is not an integer.} values of $r, \tilde r$, and possibly with  multiplicity.  A first observation is that, for odd $r$ , 
the weight of   $(r,1)$ primary is integer:
\be 
h_{2 r+ 1,1} = r \left( ( r+1)|k|  -1\right).
\ee
When combined with the right moving  with ground state with $\tilde r=1$, these are therefore candidate higher spin currents extending the Virasoro algebra to a type of W-algebra. It's an amusing feature of these models that these currents, whose spin   grows like $|k|$,
do not arise from extra gauge fields in the bulk. Instead, as we saw in the previous section, they are realized as classical solitonic solutions in the gravity sector, namely  
 spinning conical surplus solutions.

 A second important   observation is that each $(r,1)$ representation present in the spectrum should appear with a minimum multiplicity of $r-1$. This is due to the existence of  a second screening operator commuting with the Virasoro algebra which is not visible at the classical level.
 This operator is traditionally called $Q_+$ and is given by
 \be 
 Q_+ \equiv \oint dz V_{-1,1} (z) =  \oint dz : e^{-i|k| X}: (z).
 \ee
 Acting with $Q_+$ on the $(r,1)$ representation gives a $( r -1)$-fold degeneracy since $Q_+^r$ must vanish. The reason is that, as a map between Fock spaces,
 \be 
 Q_+^r: \qquad \calf_{r,s}  \to \calf_{-r,s}
 \ee 
 and $h_{-r,s} > h_{r,s}$.

In particular, the  chiral current  $\calo_{3,1}$  of lowest spin  $2 |k|-1$ is part of  a three-fold degenerate multiplet. The other currents $\calo_{2 r+1,1}$ arise from fusion (\ref{fusion})  of this basic triplet \cite{Flohr:1995ea}. The resulting W-algebra has an $SO(3)$  structure, for which $Q_+$ acts a  lowering operator, and is called the   
 'triplet algebra'. The commutation relations were worked out in \cite{Kausch:1990vg}. The vacuum module of the triplet algebra has the structure
 \be 
 \calh^W_{1,1} = \bigoplus_{r=0}^\infty \bigoplus_{i=0}^{2 r} Q_+^i \calh_{2 r+1,1}.
\ee 

More generally, we can build $2k$ irreducible representations of the triplet algebra by acting with the currents on the primary of weight  $h_{1,s}$ (which is an $SO(3)$ singlet) and on the doublet of primaries  of weight $h_{2,s}$. These have the structure  \cite{Flohr:1995ea}
\bea
\calh^W_{1,s} &=& \bigoplus_{r=0}^\infty \bigoplus_{i=0}^{2 r} Q_+^i \calh_{2 r+1,s}, \qquad s = 1, \ldots , k,\\
\calh^W_{2,s} &=& \bigoplus_{r=1}^\infty \bigoplus_{i=0}^{2 r-1} Q_+^i \calh_{2 r,s}.\label{tripletreps}
\eea 
The corresponding characters are
\bea
\chi^W_{1,s} &=& \sum_{r=0}^\infty (2r+1) \chi_{2r+1,s},\nonu
\chi^W_{2,s} &=& \sum_{r=1}^\infty 2 r \chi_{2r,s},\label{tripletchars}
\eea
where the Virasoro characters $\chi_{r,s}$ were given in (\ref{charvir}).
These can be rewritten in terms of theta functions and affine theta functions as
\bea
\chi^W_{1,s} &=&  { s \over k} { \theta_{k-s,k} \over \h} + {1 \over k} {(\pa \theta )_{k-s,k}\over \h}, \qquad s= 1, \ldots, k\nonu
\chi^W_{2,s}&=& { s \over k} { \theta_{s,k} \over \h} - {1 \over k} {(\pa \theta )_{s,k}\over \h}\label{tripchartheta}
\eea
These functions are defined  as
\bea 
\theta_{\n,k} &\equiv& \sum_{r\in \ZZ} q^{(2 k r+ \nu)^2 \over 4 k},\\
(\pa \theta)_{\n,k} &\equiv& \sum_{r\in \ZZ} (2 kr + \nu)q^{(2 k r+ \nu)^2 \over 4 k},
\eea
and satisfy 
\bea
\theta_{\n,k  } &=& \theta_{-\n,k  }= \theta_{2 k + \n,k  },\\
(\pa \theta)_{\n,k} &=& -(\pa \theta)_{-\n,k} =(\pa \theta)_{2k+\n,k}, 
\eea
so that in particular $(\pa \theta)_{0,k}=(\pa \theta)_{k,k}=0 $.

From quantizing the surplus solutions of pure gravity we found only  states belonging to the representations (\ref{tripletreps})
with $s=1$ of the triplet algebra. However, these by themselves do not combine into a modular invariant partition function, and the simplest modular invariant \cite{Flohr:1995ea} of the triplet theory involves also triplet representations with $s>1$. The resulting partition function in fact coincides with that of a non-chiral compact boson at a specific radius. We now discuss  how this partition function arises quite naturally in our free field description. 
 \subsection{A path integral argument}
 In section \ref{Secfreefield} we saw that fluctuations around a surplus solution are, in the free field variables, essentially  described by a specific winding sector of a free left- and right-moving chiral boson. This suggests  an obvious extension to a  modular invariant  theory by combining the chiral bosons into a single non-chiral compact boson for which the zero modes are coupled in the standard way. The modular invariant partition function includes additional instanton sectors coming from  solutions winding  around the Euclidean time circle.

 Concretely, we start from the Lagrangian (\ref{XLagfinal}) and  fix the gauge freedom (\ref{gaugeX},\ref{gaugeC}) by imposing  the
 gauge
 \be 
 C = \tilde C =0.
 \ee
Due to the  nonlocal  nature of the gauge transformation (\ref{gaugeX}) this step is not innocuous and  enlarges  the space of fields we integrate over. Indeed, at the perturbative level we now integrate over additional modes $a_{\pm r,n}$ in (\ref{Squadr}).

 Furthermore we replace the boundary condition  (\ref{chiralbc}) for an independent  chiral boson by one that  couples the left- and right-moving zero modes. For this purpose we define a non-chiral scalar as
 \be 
 Y(w, \bar w ) = X ( w ) + \tilde X (\bar w),
 \ee 
 and impose boundary conditions
 \bea
 Y(t_E, \f + 2 \p ) &=&  Y(t_E, \f ) + 4 \p n,\nonu
  Y(t_E + 2 \p \t_2, \f + 2 \p \t_1) &=& Y(t_E, \f ) - 4 \p m.\label{bcY}
  \eea
  Note that we allow nontrivial windings around both A- and B-cycles of the torus.
 To get a consistent variational principle for these new boundary conditions, we should also modify the boundary terms.   One can check that replacing the  boundary term (\ref{boundtermchiral}) with
 \be 
 S_{\rm bdy} = - {i |k| \over 8 \p} \int dX \wedge d\tilde X.\label{Sbdy2}
 \ee
 leads to a consistent variational principle for (\ref{bcY}). 
 
 The classical solution obeying the boundary conditions (\ref{bcY}) is
 \be 
 X = {i \over \t_2} (m- n \t ) w, \qquad \tilde X = - {i \over \t_2} (m- n \bar \t ) \bar w,
 \ee
 and the corresponding value of the  on-shell action, coming entirely from (\ref{Sbdy2}), is
 \be 
 S_E = {\p |k| \over \t_2} | m- n\t|^2.
 \ee
 Evaluating the  1-loop determinant in a standard manner \cite{Polchinski:1985zf}  and summing over all  $(n,m)$ sectors we obtain the partition function
 \bea
 Z &=& {\sqrt{|k|} \over \sqrt{\t_2} \h \bar \h }\sum_{m,n\in \ZZ} e^{-{\p |k| \over \t_2} | m- n\t|^2}\label{PFPI}\\
 &=& {1 \over \h \bar \h }\sum_{m,n\in \ZZ} q^{ (|k| n -m)^2 \over 4|k|}
 \bar  q^{ (|k| n +m)^2 \over 4|k|},\label{Zcompbos}
 \eea
 where, in the second  line, we have performed a Poisson resummation in the $m$ variable. The result is simply a free boson partition function\footnote{An alternative way to derive (\ref{PFPI}) is \cite{Henneaux:1992ig} to change variables to $Y = X + \tilde X, u = \dot X - \dot {\tilde X}$. The variable $u$ appears algebraically and upon integrating it out one obtains a standard compact boson path integral for $Y$ leading to (\ref{Zcompbos}).} at radius (in the units of \cite{Ginsparg:1987eb})
 \be 
 R = \sqrt{ |k|\over 2},
 \ee
 and, in the form (\ref{PFPI}), obviously  modular invariant.
 
 \subsection{Properties of the spectrum}
 Our proposed  modular invariant extension of the spectrum is   an obvious, though somewhat ad hoc, guess. An important sanity check is that the  partition function should decompose  into appropriate characters of the  chiral algebra at central charge $c_{|k|,1}$.  A first satisfying feature of (\ref{Zcompbos}) is that it contains, at the level of the Virasoro algebra,  only the degenerate $(r,s)$  representations. This observation goes back to \cite{DiFrancesco:1987gwq} and follows from the fact that each term in
 (\ref{Zcompbos}) can be expanded in degenerate characters using the formulas
 \begin{align}
{ q^{n^2|k|}\over \h} =& \sum_{m = n}^\infty \chi_{2m+1,|k|}, & &\nonu
{ q^{{|k|\over 4}(2 n +1)^2}\over \h} =& \sum_{m = n}^\infty \chi_{2m+2,|k|}, &  n \geq& 0,\nonu
{ q^{{1\over 4|k| }(2 |k| n +\n)^2}\over \h} =& \sum_{m = n}^\infty ( \chi_{2m+1,|k|-\n} +  \chi_{2m+2,\n}), &   n \geq& 0,\ \  1<\n<|k|,\nonu
{ q^{{1\over 4|k| }(2 |k| n -\n)^2}\over \h} =& \sum_{m = n}^\infty ( \chi_{2m,\n} +  \chi_{2m+2,|k|-\n}), & n >& 0,\ \  1<\n<|k|.\label{sumforms}
\end{align}

 A second nontrivial property \cite{Flohr:1995ea,Gaberdiel:1998ps} of the proposed the partition function (\ref{Zcompbos}) is that it also  decomposes into  characters of the much  larger triplet algebra, see (\ref{tripletchars}). To see this  we first write $Z$ in terms of theta functions (by splitting the  sum over $m$ in parts with a fixed  remainder modulo $ k$), and then use the expressions (\ref{tripchartheta}) to find
 \bea
 Z &=& \left| {\theta_{0,|k|} \over \h}  \right|^2+\left| {\theta_{|k|,|k|} \over \h}  \right|^2 +2 \sum_{\n=1}^{|k|-1} \left| {\theta_{\n,|k|} \over \h}  \right|^2\\
 &=& \left| \chi^W_{1,|k|} \right|^2 +\left| \chi^W_{2,|k|} \right|^2 + 2 \sum_{s=1}^{|k|-1}\left| \chi^W_{1,s}+ \chi^W_{2,|k|-s} \right|^2. \label{Ztripletchar}
 \eea 
It can be shown \cite{Flohr:1995ea,Flohr:2001zs} (see also \cite{Kausch:1995py,Gaberdiel:1998ps}) that (\ref{Ztripletchar}) is the minimal modular invariant combination of the characters (\ref{tripletchars}) with positive integer coefficients. The theory can be further extended by including   additional representations which  have a separately modular invariant partition function, but we will not do so here.

\begin{figure}
	\begin{center}
		\includegraphics[height=200pt]{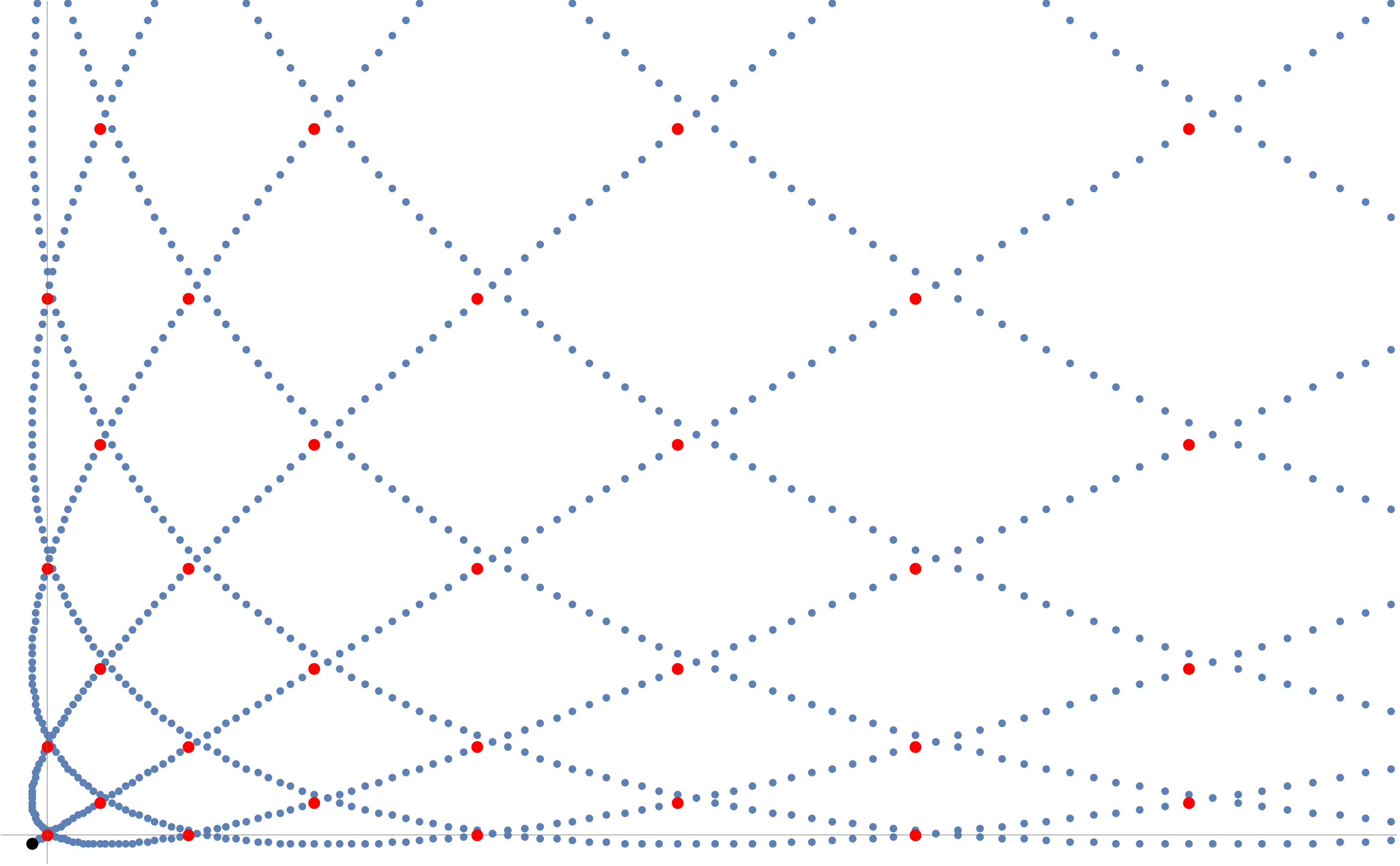}
		\put(-330,190){$\tilde h$}
			\put(-10,-10){$ h$}
	\end{center}
	\caption{Illustration of the spectrum of Virasoro primaries in the $(h, \tilde h)$ plane for $k = -12, c = -59.5$. The red dots correspond to the type $(r,1) \otimes (\tilde r,1)$ primaries which an interpretation as solitons in the Lorentzian bulk theory. The black dot is the lowest lying primary of type $(1,|k|) \otimes (1,|k|)$.}
	\label{FigSpectrum}
\end{figure}

After these consistency checks, let us examine the spectrum of Virasoro primaries in more detail. 
Using the relations (\ref{tripletchars}) (or alternatively (\ref{sumforms})), one finds that  the representations which appear are of the types
\be \begin{array}{ll} 
	(r,s ) \otimes (\tilde r, s) & {\rm for\ } r+ \tilde r\ {\rm even,}\ s = 1,\ldots k\\
	(r,s ) \otimes (\tilde r,k- s)	\qquad	 & {\rm for\ } r+ \tilde r\ {\rm odd,}\ s = 1,\ldots k .
		\end{array}
\ee
The weights of the corresponding primaries lie on parabolic curves in the $(h, \tilde h)$ plane, see Figure \ref{FigSpectrum}. 
The precise multiplicity $d(r,s ; \tilde r , \tilde s)$  of the  $(r,s ) \otimes (\tilde r, \tilde s)$ representation  can be read off from the character decomposition
\be
Z = \sum_{r,\tilde r=1}^\infty  \sum_{s,\tilde s=1}^{|k|}  d (r,s;\tilde r, \tilde s) \chi_{r,s} \bar \chi_{\tilde r, \tilde s},
\ee
and is  given by, for $|k|$ even as in the case of interest,
\begin{align}
d(r,s;\tilde r, s) =& 2 r \tilde r &  {\rm for }& \ s \notin \left\{ {|k| \over 2},|k|\right\} \ {\rm and \ }  r + \tilde r \ {\rm even}\label{dss}\nonu
d(r,s;\tilde r, |k|-s) =& 2 r \tilde r & {\rm for }& \ s\notin \left\{{|k| \over 2},|k|\right\} \  {\rm and \ }  r + \tilde r \ {\rm odd}\nonu
d\left(r,{|k| \over 2};\tilde r, {|k| \over 2}\right) =& 2 r \tilde r &&\nonu
d(r,|k|;\tilde r, |k|) =&  r \tilde r & {\rm for }& \   r + \tilde r \ {\rm even}
\end{align}
 
Let us discuss the spectrum in more detail. As we argued previously, the primaries of type $(r,1 ) \otimes (\tilde r, 1)$ have a clear Lorentzian gravity  interpretation as solitonic surplus solutions. These appear when $r + \tilde r$ is even, with multiplicity $2 r \tilde r$. 
The remaining primary states do not have a clear interpretation as smooth Lorentzian solutions but have to be included in the theory to render it consistent. Similar to the momentum sectors of a standard compact boson, they do have an interpretation as winding  sectors around the time direction in the Euclidean path integral. 

It's important to note that, while the spectrum of conformal weights is bounded below, there are states with negative weights. The state with lowest weights is the primary  $(1,|k|) \otimes  (1,|k|)$ with
\be 
h_{1,|k|} = \tilde h_{1,|k|} = {c-1\over 24}.
\ee
which appears with multiplicity one.  Since the energy scales with $c$ it is tempting to associate a classical background to this state,  which we seem to be instructed to include in the theory. This would be  the  zero mass and angular momentum limit of the BTZ black hole metric, see (\ref{BTZM0}). This solution is not smooth and the Chern-Simons gauge fields have a nontrivial holonomy of parabolic type.

It is well-known that  in CFTs where the lowest-lying state is not the conformal vacuum, the central charge $c$ appearing in the Virasoro algebra  is not the measure of the number degrees of freedom in the theory. The latter is rather measured by the effective central charge $c_{\rm eff} = c - 24 h_{\rm min}$ which is one in our case. This is the reason why the theory can have the same partition function  (\ref{Zcompbos}) as a  theory with $c=1$. The fact that  $c_{\rm eff} =1$ explains why the theory does not contain BTZ black holes  with finite horizon: while the spectrum does display Cardy growth with an exponential number of states at high level \cite{Cardy:1986ie}, the entropy does not scale with Newton's constant and is therefore too small to lead to a horizon  in the bulk description.  One can therefore  think of these models as containing only gravitational solitons and no black holes.

  \subsection{The monster of log-ness}
So far, we have discussed the $c_{|k|,1}$ models 
at the level of the partition function and its decomposition into irreducible characters. The fact that these models are in fact logarithmic CFTs is not so obvious from this point of view, though there are some indicators. A first notable feature is  that most representations appear with multiplicities   in the spectrum (cfr. the factor of 2 in (\ref{Ztripletchar})). Indeed, even the `vacuum' representation with $h_{1,1}= \tilde h_{1,1}=0$ is apparently doubly degenerate. If the spectrum is degenerated, it can happen \cite{Gurarie:1993xq} that $L_0$ (as well as the other zero-modes of the chiral algebra) are not diagonalizable but have nontrivial Jordan cells. For example $L_0$ might be represented on the two `vacuum' states as the matrix
 \be 
 L_0 = \left(\begin{array}{cc} 0 & 1\\ 0 & 0 \end{array}\right).\label{L0Jordan}
 \ee
 In this case, the two $h_{1,1}=0$ representations are part of a larger indecomposable module. 
 Theories of this type are called logarithmic CFTs, and in fact the $c_{|k|,1}$ models under consideration are  among the most studied examples (see \cite{Flohr:2001zs,Gaberdiel:2001tr,Creutzig:2013hma} for reviews). The partition function is insensitive to the presence of Jordan cells like (\ref{L0Jordan}) and it still formally decomposes into irreducible characters. The actual representations however do not decompose into irreducibles and instead form indecomposable structures.
  Another hint of the logarithmic nature of the theory is the fact that the irreducible characters  do not have nice transformation properties under modular group since the theta and affine theta functions in (\ref{tripletchars}) have different weights under the modular $S$-transformation.

  Though a thorough review of logarithmic CFTs is beyond the scope of this note, it is worth recalling how the logarithmic structure appears
   in the simplest example with $k=-2$ and $c= c_{2,1} = -2$ (which, as we recall from section \ref{Secconexc}, is strictly speaking not in our class of gravity models in  which  $k\in 4 \ZZ$). The lightest degenerate primary field is  $\calo_{2,1}$ with  $h_{2,1}=-1/8$. The four-point correlator of such fields is determined by a differential equation of second order which expresses  the decoupling of the null vector  at level two. In this special case  the roots of the characteristic equation coincide which leads to a logarithmic branch of solutions. In terms of the OPE this translates into
   \be 
   \calo_{2,1} (z) \calo_{2,1}(0) \sim z^{-{1 \over 4}} ( \calo_{1,1}( 0) + \log z \calo_{1,1}' (0)  ) + \ldots \label{logOPE}
   \ee
   Here, $\calo_{1,1}$ and $\calo_{1,1}'$ are  two primary fields of weight $h_{1,1}=0$.
   The action of $L_0$ on the corresponding states yields precisely the Jordan form of (\ref{L0Jordan}):
   \be 
   L_0 \calo_{1,1}'  = \calo_{1,1}, \qquad L_0 \calo_{1,1} =0.
   \ee
   Acting on $\calo_{1,1}'$ with the level 1 generators of the chiral algebra furthermore connects to the doublet of states $\calo_{2,1}, Q_+\calo_{2,1} $ at $h_{2,1}=1$,  schematically:
    \begin{displaymath}
   \begin{picture}(150,120)(-10,-20)
   \put(0,0){\vbox to 0pt
   	{\vss\hbox to 0pt{\hss$\bullet$\hss}\vss}}
   \put(129,0){\vbox to 0pt
   	{\vss\hbox to 0pt{\hss$\bullet$\hss}\vss}}
   \put(40,60){\vbox to 0pt
   	{\vss\hbox to 0pt{\hss$\bullet$\hss}\vss}}
   \put(89,60){\vbox to 0pt
   	{\vss\hbox to 0pt{\hss$\bullet$\hss}\vss}}
   \put(37,56){\vector(-2,-3){34}}
   \put(85,57){\vector(-3,-2){81}}
   \put(125,3){\vector(-3,2){81}}
   \put(126,5){\vector(-2,3){34}}
   \put(124,0){\vector(-1,0){119}}
   \put(-80,-3){$h_{1,1}=0$}
    \put(-80,58){$h_{2,1}=1$}
   \put(-5,-15){$\calo_{1,1}$}
   \put(124,-15){$\calo_{1,1}'$}
   \put(20,70){$Q_+\calo_{2,1}$}
   \put(80,70){$ \calo_{2,1}$}
   \end{picture}
   \end{displaymath}
   The detailed analysis of \cite{Gaberdiel:1996np}  shows that there are in this case are four consistent representations of the triplet algebra: two indecomposable ones (including the one we just sketched) and the irreducible representations which we called  $\calh^W_{1,2}, \calh^W_{2,2}$ before. The resulting CFT is rational in the sense that it has only a finite number of representations of the chiral algebra. Combining the holomorphic and antiholomorphic sectors into a local CFT leads to further nontrivial constraints as was discussed in \cite{Gaberdiel:1998ps}. The upshot for the $|k|=2$ example, which presumably generalizes to arbitrary $|k|$, is that the entire  last term containing the sum in (\ref{Ztripletchar}) forms the character of a single indecomposable representation.   
 
 \section{Discussion}
 In this work we made a case for a holographic dual interpretation of the non-unitary $c_{p,1}$ logarithmic models at large $p$. 
 It is interesting to contrast the proposal for a holographic duality involving a specific CFT with recently studied  instances of
 `imprecise' holography where a gravity-like theory in the bulk is described by an  average over an ensemble of CFTs. In the recent example of \cite{Afkhami-Jeddi:2020ezh,Maloney:2020nni}, as  in the original work on pure gravity \cite{Maloney:2007ud}  the partition function is obtained by starting from the Virasoro character of the lowest energy state in the theory and  performing a (regularized)  sum 
 over modular images. This leads, under natural assumptions, to a continuous spectrum characteristic of an ensemble average of CFTs. However,  in theories with $c_{\rm eff}=1$, the procedure of modular averaging becomes much more subtle. Indeed, 
 the standard expression for the modular average becomes ill-defined (see e.g. (2.3) in \cite{Benjamin:2020mfz}) in this case. Both  the models we considered here and the proposal for a gravity dual  to Liouville theory in \cite{Li:2019mwb} have $c_{\rm eff}=1$, and this seems to be how these examples bypass the assumptions leading to an ensemble-averaged partition function. It would be interesting to understand how this works for other examples of precise  holography such as  tensionless string theory on AdS$_3$ \cite{Eberhardt:2018ouy}.
 
We end by listing some remaining confusions and possible generalizations.
\begin{itemize}
\item While, as we have argued, the $(r,1 ) \otimes (\tilde r, 1)$ representations have a clear Lorentzian bulk  interpretation, the same can not be said of the  other representations which were needed to fill out a modular invariant spectrum. These arose from winding  sectors around the time direction in the Euclidean gravitational path integral. It is at present not clear if these represent fluctuations of matter fields that we should add to the theory (as was the case in the example of \cite{Perlmutter:2012ds}, where similar representations came from a Vasiliev-like scalar field \cite{Prokushkin:1998bq}), or whether they should be seen as somewhat exotic boundary graviton sectors. 
\item	It is somewhat unsatisfying that, in the current work,  the logarithmic character of the dual theory was not directly visible in the bulk, rather we were led to it by considering the modular invariant completion of the spectrum. It would be interesting to see a logarithmic branch (\ref{logOPE}) appearing in an amplitude computed in the bulk. This should happen for example in the four-point amplitude of the lowest lying primary in the theory, which in the bulk is represented as an $M=J=0$ BTZ metric.
\item It would be interesting to understand better if the standard nonunitary minimal models with $c_{p,p'}$ for $p>p'>1$  also have  a gravity-like  dual interpretation at large negative central charge. 
One would expect that the $(r,1)$ representations still  have an interpretation as surplus solutions. However, the structure of the degenerate representations and of the partition function is more complicated in this case, and perhaps  a more sophisticated version of our arguments could lead to these models.
 \item It is expected that our observations can be extended to higher spin theories based on an $SL(N,\RR) \times SL(N,\RR)$ Chern-Simons theory in the bulk.  It would be interesting to understand if one obtains, following the procedure of \cite{Cotler:2018zff} and section \ref{Secconexc} an analog of the geometric action for W$_N$ coadjoint orbits. Also, the argument of \cite{Campoleoni:2017xyl} picks out the value of the central charge
 \be 
 c^N_{|k|,1} =(N-1)\left( 1 - N (N+1) {( |k|- 1 )^2\over |k|}\right)
 \ee
 which for integral $k$ is a  limiting case of the central charge for W$_N$ minimal models.
 	\end{itemize}

 \section*{Acknowledgements}
 I would like to thank Gideon Vos for useful discussions and Dio Anninos, Andrea Campoleoni,  Tom\'{a}\v{s} Proch\'{a}zka and Stefan Fredenhagen for valuable comments on the draft. This work was supported
 by the Grant Agency of the Czech Republic under the grant EXPRO 20-25775X.
 
\begin{appendix}
\section{Variational principle and boundary term}\label{Appbdyterm}
In this section we briefly review, following \cite{Henneaux:1987hz}, the boundary terms which have to be added to (\ref{XLagfinal}) to obtain a well-defined variational principle.  This requires that we specify, in addition to the quasi-periodic boundary conditions along the $\f$-circle, boundary conditions at initial and final times (say, $t_1$ and $t_2$) and  possibly add appropriate boundary terms. Naively, the action is stationary on solutions of the equations of motion 
(\ref{eomfinal}) if we keep $X$ fixed at both endpoints:
\be 
\d X (t_1,\f) = \d X (t_2,\f)=0.
\ee
However, as stressed in \cite{Henneaux:1987hz}, this is not correct as it  imposes two independent conditions while $X$ only obeys a first-order  equation in time. As was argued there, a natural boundary condition to  impose is 
\be 
\d X (t_1 , \f) = - \d X (t_2, \f).\label{chiralbc}
\ee
One checks that the action is stationary on solutions obeying this boundary conditions if we add the boundary term
\be 
S_{\rm bdy} = - {|k| \over 16 \p} \int_0^{2 \p} d \f \left(  X(t_2, \f) - X(t_1,\f)\right) \left(  X'(t_2, \f) + X'(t_1,\f)\right).\label{boundtermchiral}
\ee

The presence of this boundary term also solves a puzzle about gauge invariance of the on-shell action. Using the shift transformation (\ref{shiftX}) we can represent the surplus solutions either as
\be 
X = r x_+ \qquad {\rm or } \qquad X = r \f + c_0.\label{2forms}
\ee
Without the boundary term, the first form would lead to vanishing on-shell action while the second would give
\be
S_{\rm on-shell}= - {|k| n^2 \over 4} (t_2-t_1)\label{Sos}
\ee
This is the physically expected answer as it equals minus the energy of the solution times the time interval. Adding the boundary however term both forms in (\ref{2forms}) give the correct answer (\ref{Sos}). The boundary term is required for on-shell gauge invariance because the gauge transformation linking the two solutions does not vanish at the endpoints $t_1$ and $t_2$, though it does respect the boundary condition (\ref{chiralbc}) for an appropriate choice of the constant $c_0$ in  (\ref{2forms}).

\end{appendix}
\bibliographystyle{ytphys}
\bibliography{ref}
\end{document}